\documentclass[pre,aps,showpacs,article]{revtex4}
\usepackage{graphicx,amsmath,bm}
\usepackage{amsfonts}
\usepackage{amssymb}
\usepackage{multirow} %for multi-row in tables

\begin{document}

\title{Near linear time algorithm to detect community structures in large-scale networks}
\author{$^1$Usha Nandini Raghavan, $^2$R\'eka Albert and $^1$Soundar Kumara}
\affiliation{$^1$Department of Industrial Engineering, The
Pennsylvania State University, University Park, Pennsylvania, 16802,
USA\\
$^2$Department of Physics, The Pennsylvania State University,
University Park, Pennsylvania, 16802, USA}

\begin{abstract}
Community detection and analysis is an important methodology for
understanding the organization of various real-world networks and
has applications in problems as diverse as consensus formation in
social communities or the identification of functional modules in
biochemical networks. Currently used algorithms that identify the
community structures in large-scale real-world networks require
\emph{a priori} information such as the number and sizes of
communities or are computationally expensive.  In this paper we
investigate a simple label propagation algorithm that uses the
network structure alone as its guide and requires neither
optimization of a pre-defined objective function nor prior
information about the communities. In our algorithm every node is
initialized with a unique label and at every step each node adopts
the label that most of its neighbors currently have. In this
iterative process densely connected groups of nodes form a consensus
on a unique label to form communities. We validate the algorithm by
applying it to networks whose community structures are known.  We
also demonstrate that the algorithm takes an almost linear time and
hence it is computationally less expensive than what was possible so
far.
\end{abstract}
\pacs{89.75.Fb, 89.75.Hc, 87.23.Ge, 02.10.Ox}

\maketitle

\section{Introduction}
A wide variety of complex systems can be represented as networks.
For example, the World Wide Web is a network of webpages
interconnected by hyperlinks; social networks are represented by
people as nodes and their relationships by edges; and biological
networks are usually represented by bio-chemical molecules as nodes
and the reactions between them by edges. Most of the research in the
recent past focused on understanding the evolution and organization
of such networks and the effect of network topology on the dynamics
and behaviors of the system
\cite{Albert02,Albert99,Barabasi99,Newman03}. Finding community
structures in networks is another step towards understanding the
complex systems they represent.

A community in a network is a group of nodes that are similar to
each other and dissimilar from the rest of the network. It is
usually thought of as a group where nodes are densely
inter-connected and sparsely connected to other parts of the network
\cite{Girvan02,Newman03,Wasserman94}. There is no universally
accepted definition for a community, but it is well known that most
real-world networks display community structures. There has been a
lot of effort recently in defining, detecting and identifying
communities in real-world networks
\cite{Danon06,Eckmann02,Flake00,Girvan02,Guimera05,Gustafsson06,Hastings06,Newman04a,Palla05,Radicchi04}.
The goal of a community detection algorithm is to find groups of
nodes of interest in a given network. For example, a community in
the WWW network indicates a similarity among nodes in the group.
Hence, if we know the information provided by a small number of
webpages, then it can be extrapolated to other webpages in the same
community. Communities in social networks can provide insights about
common characteristics or beliefs among people that makes them
different from other communities. In bio-molecular interaction
networks, segregating nodes into functional modules can help
identify the roles or functions of individual molecules
\cite{Guimera05}. Further, in many large-scale real-world networks,
communities can have distinct properties which are lost in their
combined analysis \cite{Albert02}.

Community detection is similar to the well studied network
partitioning problems \cite{Karger00,Kernighan70,Fiduccia82}. The
\emph{network partitioning} problem is in general defined as the
partitioning of a network into $c$ (a fixed constant) groups of
approximately equal sizes, minimizing the number of edges between
groups. This problem is NP-hard and efficient heuristic methods have
been developed over years to solve the problem
\cite{Fiduccia82,Hendrickson95,Karger00,Kernighan70,Stoer97}. Much
of this work is motivated by engineering applications including very
large scale integrated (VLSI) circuit layout design and mapping of
parallel computations. Thompson \cite{Thompson79} showed that one of
the important factors affecting the minimum layout area of a given
circuit in a chip is its bisection width. Also, to enhance the
performance of a computational algorithm, where nodes represent
computations and edges represent communications, the nodes are
divided equally among the processors so that the communications
between them are minimized.

The goal of a network partitioning algorithm is to divide any given
network into approximately equal size groups irrespective of node
similarities. Community detection on the other hand finds groups
that either have an inherent or an externally specified notion of
similarity among nodes within groups. Furthermore, the number of
communities in a network and their sizes are not known beforehand
and they are established by the community detection algorithm.

Many algorithms have been proposed to find community structures in
networks. Hierarchical methods divide networks into communities,
successively, based on a dis-similarity measure, leading to a series
of partitions from the entire network to singleton communities
\cite{Girvan02,Radicchi04}. Similarly one can also successively
group together smaller communities based on a similarity measure
leading again to a series of partitions \cite{Newman04b,Pons05}. Due
to the wide range of partitions, structural indices that measure the
strength of community structures are used in determining the most
relevant ones. Simulation based methods are also often used to find
partitions with a strong community structure
\cite{Duch05,Guimera05}. Spectral \cite{Kernighan70,Newman06} and
flow maximization (cut minimization) methods \cite{Flake00,Wu04}
have been successfully used in dividing networks into two or more
communities.

In this paper, we propose a localized community detection algorithm
based on label propagation. Each node is initialized with a unique
label and at every iteration of the algorithm, each node adopts a
label that a maximum number of its neighbors have, with ties broken
uniformly randomly. As the labels propagate through the network in
this manner, densely connected groups of nodes form a consensus on
their labels. At the end of the algorithm, nodes having the same
labels are grouped together as communities. As we will show, the
advantage of this algorithm over the other methods is its simplicity
and time efficiency. The algorithm uses the network structure to
guide its progress and does not optimize any specific chosen measure
of community strengths. Furthermore, the number of communities and
their sizes are not known a priori and are determined at the end of
the algorithm. We will show that the community structures obtained
by applying the algorithm on previously considered networks, such as
Zachary's karate club friendship network and the US college football
network, are in agreement with the actual communities present in
these networks.

\section{Definitions and previous work}
As mentioned earlier, there is no unique definition of a community.
One of the simplest definitions of a community is a clique, that is,
a group of nodes where there is an edge between every pair of nodes.
Cliques capture the intuitive notion of a community
\cite{Wasserman94} where every node is related to every other node
and hence have strong similarities with each other. An extension of
this definition was used by Palla et al in \cite{Palla05}, who
define a community as a chain of adjacent cliques. They define two
$k$-cliques (cliques on $k$ nodes) to be adjacent if they share
$k-1$ nodes. These definitions are strict in the sense that the
absence of even one edge implies that a clique (and hence the
community) no longer exists. $k$-clans and $k$-clubs are a more
relaxed definitions while still maintaining a high density of edges
within communities \cite{Palla05}. A group of nodes is said to form
a $k$-clan if the shortest path length between any pair of nodes, or
the diameter of the group, is at most $k$. Here the shortest path
only uses the nodes within the group. A $k$-club is defined
similarly, except that the subnetwork induced by the group of nodes
is a maximal subgraph of diameter $k$ in the network.

Definitions based on degrees {(number of edges)} of nodes within the
group relative to their degrees outside the group were given by
Radicchi et al \cite{Radicchi04}. If $d_i^{in}$ and $d_i^{out}$ are
the degrees of node $i$ within and outside of its group $U$, then
$U$ is said to form a \emph{strong} community if
$d_i^{in}>d_i^{out},\text{ }\forall i \in U$. If $\sum_{i\in
U}d_i^{in}>\sum_{i\in U}d_i^{out}$, then $U$ is a community in the
\emph{weak} sense. Other definitions based on degrees of nodes can
be found in \cite{Wasserman94}.

There can exist many different partitions of nodes in the network
that satisfy a given definition of community. In most cases
\cite{Bagrow05,costa04,Newman03,Newman04b,Wu04}, the groups of nodes
found by a community detection algorithm are assumed to be
communities irrespective of whether they satisfy a specific
definition or not. To find the best community structures among them
we need a measure that can quantify the strength of a community
obtained. One of the ways to measure the strength of a community is
by comparing the density of edges observed within the community with
the density of edges in the network as a whole \cite{Wasserman94}.
If the number of edges observed within a community $U$ is $e_U$,
then under the assumption that the edges in the network are
uniformly distributed among pairs of nodes, we can calculate the
probability $P$ that the expected number of edges within $U$ is
larger than $e_U$. If $P$ is small, then the observed density in the
community is greater than the expected value. A similar definition
was recently adopted by Newman \cite{Newman04a}, where the
comparison is between the observed density of edges within
communities and the expected density of edges within the same
communities in randomized networks that nevertheless maintain every
node's degree. This was termed the \emph{modularity measure Q},
where $Q = \sum_i (e_{ii} - a_i^2),\text{  }\forall i$. $e_{ii}$ is
the observed fraction of edges within group $i$ and $a_i^2$ is the
expected fraction of edges within the same group $i$. Note that if
$e_{ij}$ is the fraction of edges in the network that run between
group $i$ and group $j$, then $a_i = \sum_j e_{ij}$. $Q=0$ implies
that the density of edges within groups in a given partition is no
more than what would be expected by a random chance. $Q$ closer to 1
indicates stronger community structures.

Given a network with $n$ nodes and $m$ edges $N(n,m)$, any community
detection algorithm finds subgroups of nodes. Let $C_1,C_2,...,C_p$
be the communities found. In most algorithms, the communities found
satisfy the following constraints
\begin{enumerate}
\item $C_i\cap C_j = \emptyset$ for $i\neq j$ and
\item $\bigcup_i {C_i}$ spans the node set in $N$
\end{enumerate}
A notable exception is Palla et al \cite{Palla05} who define
communities as a chain of adjacent $k$-cliques and allow community
overlaps. It takes exponential time to find all such communities in
the network. They use these sets to study the overlapping structure
of communities in social and biological networks. By forming another
network where a community is represented by a node and edges between
nodes indicate the presence of overlap, they show that such networks
are also heterogeneous (fat-tailed) in their node degree
distributions. Furthermore, if a community has overlapping regions
with two other communities, then the neighboring communities are
also highly likely to overlap.

\begin{figure*}
\begin{center}
  \includegraphics[width=8cm]{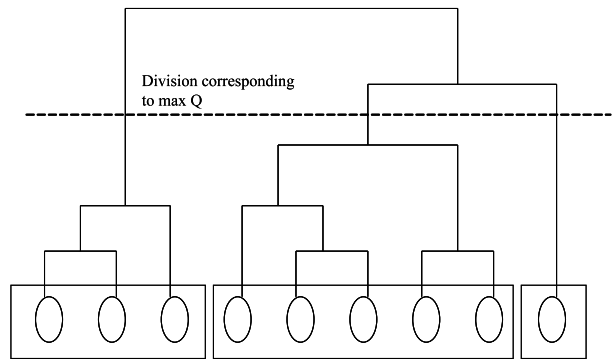}\\
  \caption{An illustration of a dendrogram which is a tree representation of the
  order in which nodes are segregated into different groups or communities.}\label{dendogram}
\end{center}
\end{figure*}

The number of different partitions of a network $N(n,m)$ into just
two disjoint subsets is $2^n$ and increases exponentially with $n$.
Hence we need a quick way to find only relevant partitions. Girvan
and Newman \cite{Girvan02} proposed a divisive algorithm based on
the concept of \emph{edge betweenness centrality }, that is, the
number of shortest paths among all pairs of nodes in the network
passing through that edge. The main idea here is that edges that run
between communities have high betweenness values than those that lie
within communities. By successively recalculating and removing edges
with highest betweenness values, the network breaks down into
disjoint connected components. The algorithm continues until all
edges are removed from the network. Each step of the algorithm takes
$O(mn)$ time and since there are $m$ edges to be removed, the worst
case running time is $O(m^2n)$. As the algorithm proceeds {one can}
construct a dendrogram (see figure \ref{dendogram}) depicting the
breaking down of the network into disjoint connected components.
Hence for any given $h$ such that $1\leq h \leq n$, at most one
partition of the network into $h$ disjoint subgroups {is found. All}
such partitions in the dendrogram {are depicted}, irrespective of
whether or not the subgroups in each partition {represent } a
community. Radicchi et al \cite{Radicchi04} propose another divisive
algorithm where the dendrograms are modified to reflect only those
groups that satisfy a specific definition of a community. Further,
instead of edge betweenness centrality, they use a local measure
called \emph{edge clustering coefficient} as a criterion for
removing edges. The edge clustering coefficient is defined as the
fraction of number of triangles a given edge participates in, to the
total number of possible such triangles. The clustering coefficient
of an edge is expected to be the least for those running between
communities and hence the algorithm proceeds by removing edges with
low clustering coefficients. The total running time of this divisive
algorithm is $O(\frac{m^4}{n^2})$.

Similarly one can also define a topological similarity between nodes
and perform an agglomerative hierarchical clustering
\cite{Newman04c,Pons05}. In this case, we begin with nodes in $n$
different communities and group together communities that are the
most similar. Newman \cite{Newman04b} proposed an amalgamation
method (similar to agglomerative methods) using the modularity
measure Q, where at each step those two communities are grouped
together that give rise to the maximum increase or smallest decrease
in Q. This process can also be represented as a dendrogram and one
can cut across the dendrogram to find the partition corresponding to
the maximum value of Q (see figure \ref{dendogram}). At each step of
the algorithm {one compares} at most $m$ pairs of groups and
{requires} at most $O(n)$ time to update the $Q$ value. The
algorithm continues until all the $n$ nodes are in one group and
hence the worst case running time of the algorithm is $O(n(m+n))$.
The algorithm of Clauset et al \cite{Clauset04} is an adaptation of
this agglomerative hierarchical method, but uses a clever data
structure to store and retrieve information required to update $Q$.
In effect, they reduce the time complexity of the algorithm to
{$O(md \text{ log }n)$,} where $d$ is the depth of the dendrogram
obtained. In networks that have a hierarchical structure with
communities at many scales, {$d\sim \text{log }n$.} There have also
been other heuristic and simulation based methods that find
partitions of a given network maximizing the modularity measure $Q$
\cite{Duch05,Guimera05}.

Label flooding algorithms have also been used in detecting
communities in networks \cite{Bagrow05,costa04}. In \cite{Bagrow05},
the authors propose a local community detection method where a node
is initialized with a label which then propagates step by step via
the neighbors until it reaches the end of the community, where the
number of edges proceeding outward from the community drops below a
threshold value. After finding the local communities at all nodes in
the network, an $n\times n$ matrix is formed, where the
$ij^{\text{th}}$ entry is 1 if node $j$ belongs to the community
started from $i$ and 0 otherwise. The rows of the matrix are then
rearranged such that the similar ones are closer to each other.
Then, starting from the first row they successively include all the
rows into a community until the distance between two successive rows
is large and above a threshold value. After this a new community is
formed and the process is continued. Forming the rows of the matrix
and rearranging them requires $O(n^3)$ time and hence the algorithm
is time-consuming.

Wu and Huberman \cite{Wu04} propose a linear time ($O(m+n)$)
algorithm that can divide a given network into two communities.
Suppose that {one} can find two nodes ($x$ and $y$) that belong to
two different communities, then they are initialized with values 1
and 0 respectively. All other nodes are initialized with value 0.
Then at each step of the algorithm, all nodes (except $x$ and $y$)
update their values as follows. If $z_1,z_2,...,z_k$ are neighbors
of a node $z$, then the value $V_z$ is updated as $\frac{V_{z_1} +
V_{z_2}+...+V_{z_k}}{k}$. This process continues until convergence.
The authors show that the iterative procedure converges to a unique
value, and the convergence of the algorithm does not depend on the
size $n$ of the network. Once the required convergence is obtained,
the values are sorted between 0 and 1. Going through the spectrum of
values in descending order, there will be a sudden drop at the
border of two communities. This gap is used in identifying the two
communities in the network. A similar approach was used by Flake et
al \cite{Flake00} to find the communities in the WWW network. Here,
given a small set of nodes (source nodes), they form a network of
webpages that are within a bounded distance from the sources. Then
by designating (or artificially introducing) sink nodes, they solve
for the maximum flow from the sources to the sinks. In doing so one
can then find the \emph{minimum cut} corresponding to the maximum
flow. The connected component of the network containing the source
nodes after the removal of the cut set is then the required
community.

Spectral bisection methods \cite{Newman06} have been used
extensively to divide a network into two groups so that the number
of edges between groups is minimized. Eigenvectors of the Laplacian
matrix ($L$) of a given network are used in the bisection process.
It can be shown that $L$ has only real non-negative eigenvalues
($0\leq \lambda_1 \leq \lambda_2 \leq ...\leq \lambda_n$) and
minimizing the number of edges between groups is same as minimizing
the positive linear combination $M = \sum_i s_i^2 \lambda_i$, where
$s_i = u_i^Tz$ and $u_i$ is the eigenvector of L corresponding to
$\lambda_i$. $z$ is the decision vector whose $i^{\text{th}}$ entry
can be either 1 or -1 denoting which of the two groups node $i$
belongs to. To minimize $M$, $z$ is chosen as parallel as possible
to the eigenvector corresponding to the second smallest eigenvalue
(The smallest eigenvalue is $0$ and choosing $z$ parallel to the
corresponding eigenvector gives a trivial solution). This bisection
method has been extended to finding communities in networks that
{maximize} the modularity measure $Q$ \cite{Newman06}. $Q$ can be
written as a positive linear combination of eigenvalues of the
matrix $B$, where $B$ is defined as difference of the two matrices
$A$ and $P$. $A_{ij}$ is the observed number of edges between nodes
$i$ and $j$ and $P_{ij}$ is the expected number of edges between $i$
and $j$ if the edges fall randomly between nodes, while maintaining
the degree of each node. Since $Q$ has to be maximized, $z$ is
chosen as parallel as possible to the eigenvector corresponding to
the largest eigenvalue.

Since many real-world complex networks are large in size, time
efficiency of the community detection algorithm is an important
consideration. When no \emph{a priori} information is available
about the likely communities in a given network, finding partitions
that optimize a chosen measure of community strength is normally
used. Our goal in this paper is to develop a simple time-efficient
algorithm that requires no prior information (such as number, sizes
or central nodes of the communities) and uses only the network
structure to guide the community detection. The proposed {mechanism
for} such an algorithm which does not optimize any specific measure
or function is detailed in the following section.

\section{Community detection using label propagation}

\begin{figure*}
\begin{center}
  \includegraphics[width=8cm]{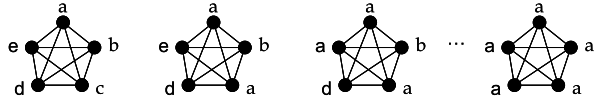}\\
  \caption{Nodes are updated one by one as we move from left to right. Due to a
  high density of edges (highest possible in this case), all nodes acquire the
  same label.}\label{consensus}
\end{center}
\end{figure*}

The main idea behind our label propagation algorithm is the
following. Suppose that a node $x$ has neighbors $x_1,x_2,...,x_k$
and that each neighbor carries a label denoting the community to
which they belong to. Then $x$ determines its community based on the
labels of its neighbors. We assume that each node in the network
chooses to join the community to which the maximum number of its
neighbors belong to, with ties broken uniformly randomly. We
initialize every node with unique labels and let the labels
propagate through the network. As the labels propagate, densely
connected groups of nodes quickly reach a consensus on a unique
label (see figure \ref{consensus}). When many such dense (consensus)
groups are created throughout the network, they continue to expand
outwards until it is possible to do so. At the end of the
propagation process, nodes having the same labels are grouped
together as one community.

\begin{figure*}
\begin{center}
  \includegraphics[width=8cm]{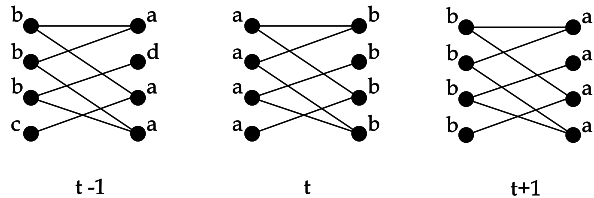}\\
  \caption{An example of a bi-partite network in which the label sets of the two
  parts are disjoint. In this case, due to the choices made by the nodes at step
  $t$, the labels on the nodes oscillate between $a$ and $b$. }\label{oscillations}
\end{center}
\end{figure*}

We perform this process iteratively, where at every step, each node
updates its label based on the labels of its neighbors. The updating
process can either be synchronous or asynchronous. In synchronous
updating, node $x$ at the $t^{\text{th}}$ iteration updates its
label based on the labels of its neighbors at iteration $t-1$.
Hence, $C_x(t) = f(C_{x_1}(t-1),...,C_{x_k}(t-1))$, where $c_x(t)$
is the label {of} node $x$ at time $t$. The problem however is that
subgraphs in the network that are bi-partite or nearly bi-partite in
structure lead to oscillations of labels (see figure
\ref{oscillations}). This is especially true in cases where
communities take the form of a star graph. Hence we use asynchronous
updating where $C_x(t) =
f(C_{x_{i1}}(t),...,C_{x_{im}}(t),C_{x_{i(m+1)}}(t-1),...,C_{x_{ik}}(t-1))$
and $x_{i1},...,x_{im}$ are neighbors of $x$ that have already been
updated in the current iteration while $x_{i(m+1)},...,x_{ik}$ are
neighbors that are not yet updated in the current iteration. The
order in which all the $n$ nodes in the network are updated at each
iteration is chosen randomly. Note that while we have $n$ different
labels at the beginning of the algorithm, the number of labels
reduces over iterations, resulting in only as many unique labels as
there are communities.

Ideally the iterative process should continue until no node in the
network changes its label. However, there could be nodes in the
network that have an equal maximum number of neighbors in two or
more communities. Since we break ties randomly among the possible
candidates, the labels on such nodes could change over iterations
even if the labels of their neighbors remain constant. Hence we
perform the iterative process until every node in the network has a
label to which the maximum number of its neighbors belong to. By
doing so we obtain a partition of the network into disjoint
communities, where every node has at least as many neighbors within
its community as it has with any other community. If $C_1,...,C_p$
are the labels that are currently active in the network and
$d_i^{C_j}$ is the number of neighbors node $i$ has with nodes of
label $C_j$, then the algorithm is stopped when for every node $i$,
$$\text{If } i \text{ has label } C_m \text{ then } d_i^{C_m}\geq d_i^{C_j}\text{
}\forall j$$ At the end of the iterative process nodes with {the}
same label are grouped together as communities. Our stop criterion
characterizing the obtained communities is similar (but not
identical) to the definition of \emph{strong communities} proposed
by Radicchi et al \cite{Radicchi04}. While strong communities
require each node to have strictly more neighbors within its
community than outside, the communities obtained by the label
propagation process require each node to have at least as many
neighbors within its community as it has with each of the other
communities. We can describe our proposed label propagation
algorithm in the following steps.
\begin{enumerate}
\item Initialize the labels at all nodes in the network. For a given node $x$, $C_x(0) = x$.
\item Set $t = 1$.
\item Arrange the nodes in the network in a random order and set it
to $X$.
\item For each $x\in X$ chosen in that specific order, let $C_x(t) =
f(C_{x_{i1}}(t),...,C_{x_{im}}(t),C_{x_{i(m+1)}}(t-1),...,C_{x_{ik}}(t-1))$.
$f$ here returns the label occurring with the highest frequency
among neighbors and ties are broken uniformly randomly.
\item If every node has a label that the maximum number of their
neighbors have, then stop the algorithm. Else, set $t = t+1$ and go
to (3).
\end{enumerate}
Since we begin the algorithm with each node carrying a unique label,
the first few iterations result in various small pockets (dense
regions) of nodes forming a consensus (acquiring the same label).
These consensus groups then gain momentum and try to acquire more
nodes to strengthen the group. However, when a consensus group
reaches the border of another consensus group, they start to compete
for members. The within-group interactions of the nodes can
counteract the pressures from outside if there are less
between-group edges than within-group edges. The algorithm
converges, and the final communities are identified, when a global
consensus among groups is reached. Note that even though the network
as one single community satisfies the stop criterion, this process
of group formation and competition discourages all nodes from
acquiring the same label in case of heterogeneous networks with an
underlying community structure. In case of homogeneous networks such
as Erd\H{o}s - R\'enyi random graphs \cite{Bollobas85} that do not
have community structures, the label propagation algorithm
identifies the giant connected component of these graphs as a single
community.

\begin{figure*}
  \includegraphics[width=6.6cm]{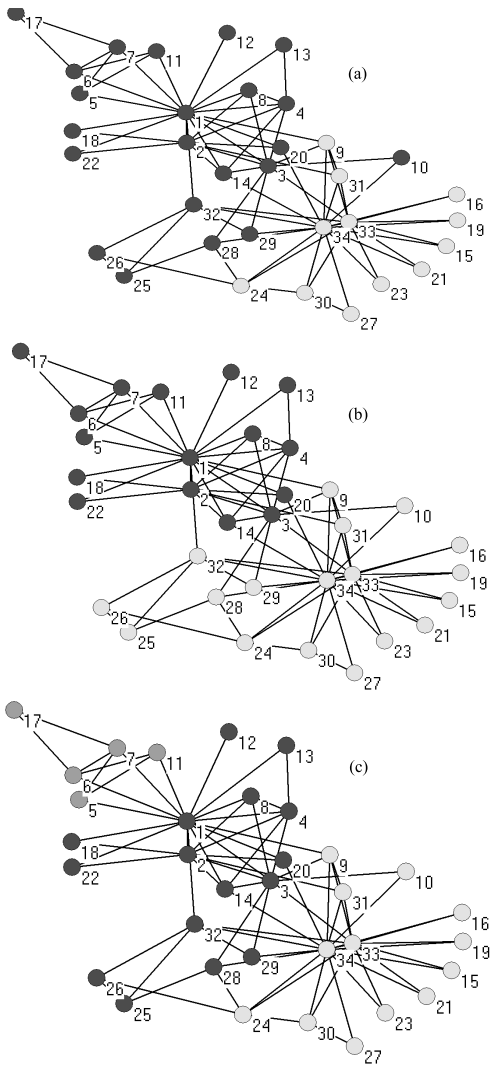}\\
  \caption{$(a),(b)$ and $(c)$ are three different community structures
  identified by the algorithm on Zachary's karate club network. The communities
  can be identified by their shades of grey colors.}\label{karatefootball}
\end{figure*}

\begin{figure*}
\begin{center}
  \includegraphics[width=10.8cm]{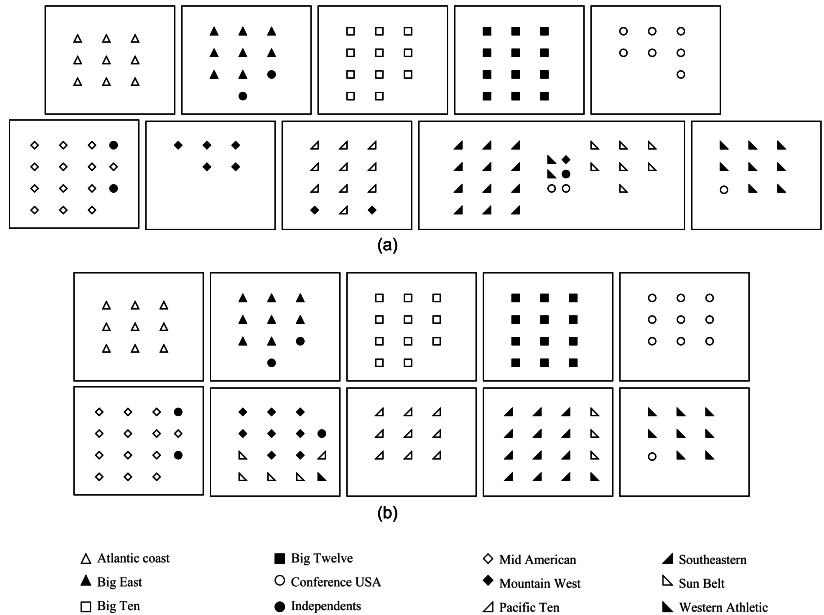}\\
  \caption{The grouping of US college football teams into conferences are shown
  in $(a)$ and $(b)$. Each solution ($(a)$ and $(b)$) is an aggregate of five
  different solutions obtained by applying the algorithm on the college football
  network.}\label{football}
\end{center}
\end{figure*}

Our stop criterion is only a condition and not a measure that is
being maximized or minimized. Consequently there is no unique
solution and more than one distinct partition of a network into
groups satisfies the stop criterion (see figures
\ref{karatefootball} and \ref{football}). Since the algorithm breaks
ties uniformly randomly, early on in the iterative process when
possibilities of ties are high, a node may vote in favor of a
randomly chosen community. As a result, multiple community
structures are reachable from the same initial condition.

If we know the set of nodes in the network that are likely to act as
centers of attraction for their respective communities, then it
would be sufficient to initialize such nodes with unique labels,
leaving the remaining nodes unlabeled. In this case when we apply
the proposed algorithm the unlabeled nodes will have a tendency to
acquire labels from their closest attractor and join that community.
Also, restricting the set of nodes initialized with labels will
reduce the range of possible solutions that the algorithm can
produce. Since it is generally difficult to identify nodes that are
central to a community before identifying the community itself, here
we give all nodes equal importance at the beginning of the algorithm
and provide them each with unique labels.

We apply our algorithm to the following networks. The first one is
Zachary's karate club network which is a network of friendship among
34 members of a karate club \cite{Zachary77}. Over a period of time
the club split into two factions due to leadership issues and each
member joined one of the two factions. The second network that we
consider is the US college football network that consists of 115
college teams represented as nodes and has edges between teams that
played each other during the regular season in the year 2000
\cite{Girvan02}. The teams are divided into conferences
(communities) and each team plays more games within its own
conference than inter-conference games. Next is the co-authorship
network of 16726 scientists who have posted preprints on the
condensed matter archive at www.arxiv.org; the edges connect
scientists who co-authored a paper \cite{Newman01}. It has been
shown that communities in co-authorship networks are made up by
researchers working in the same field or are research groups
\cite{Newman04b}. Along similar lines one can expect an actor
collaboration network to have communities containing actors of a
similar genre. Here we consider an actor collaboration network of
374511 nodes and edges running between actors who have acted in at
least one movie together \cite{Barabasi99}. We also consider a
protein-protein interaction network \cite{Jeong00} consisting of
2115 nodes. The communities are likely to reflect functional
groupings of this network. And finally we consider a subset of the
world wide web (WWW) consisting of 325729 webpages within the nd.edu
domain and hyperlinks interconnecting them \cite{Albert99}.
Communities here are expected to be groups of pages on similar
topics.

\subsection{Multiple community structures}
\begin{figure*}
\begin{center}
  \includegraphics[width=15cm]{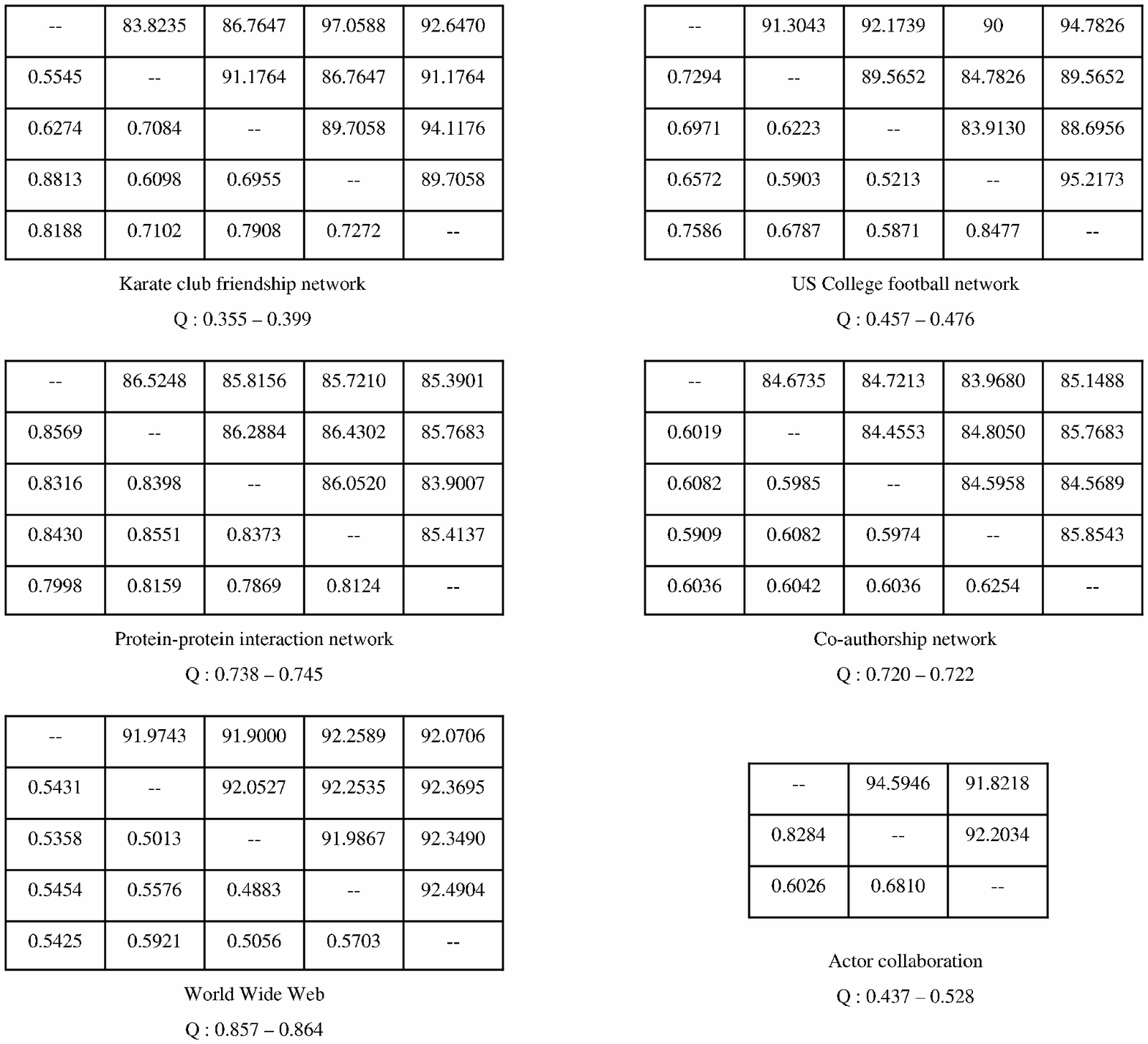}\\
  \caption{Similarities between five different solutions obtained for each
  network is tabulated. An entry in the $i^{\text{th}}$ row and $j^{\text{th}}$
  column in the lower triangle of each of the table is the Jaccard's similarity
  index for solutions $i$ and $j$ of the corresponding network. Entries in the
  $i^{\text{th}}$ row and $j^{\text{th}}$ column in the upper-triangle of the
  tables are the values of the measure $f_{same}$ for solutions $i$ and $j$ in
  the respective networks. The range of modularity values $Q$ obtained for the
  five different solutions is also given for each network.}\label{solutions}
\end{center}
\end{figure*}

Figure \ref{karatefootball} shows three different solutions obtained
for the Zachary's karate club network and figure \ref{football}
shows two different solutions obtained for the US college football
network. We will show that even though we obtain different solutions
(community structure), they are similar to each other. To find the
percentage of nodes classified in the same group in two different
solutions, we form a matrix $M$, where $M_{ij}$ is the number of
nodes common to community $i$ in one solution and community $j$ in
the other solution. Then we calculate $f_{same} =
\frac{1}{2}(\sum_i\text{max}_j\{M_{ij}\}+\sum_j\text{max}_i\{M_{ij}\})\frac{100}{n}$.
Given a network whose communities are already known, a community
detection algorithm is commonly evaluated based on the percentage
(or number) of nodes that are grouped into the correct communities
\cite{Newman04b,Wu04}. $f_{\text{same}}$ is similar, where by fixing
one solution we evaluate how close the other solution is to the
fixed one and vice versa. While $f_{same}$ can identify how close
one solution is to another, it is however not sensitive to the
seriousness of errors. For example, when few nodes from several
different communities in one solution are fused together as a single
community in another solution, the value of $f_{same}$ does not
change much. Hence we also use Jaccard's index which has been shown
to be more sensitive to such differences between solutions
\cite{Milligan85}. If $a$ stands for the pairs of nodes that are
classified in the same community in both solutions, $b$ for pairs of
nodes that are in the same community in the first solution and
different in the second and $c$ vice-versa, then Jaccard's index is
defined as $\frac{a}{a+b+c}$. It takes values between 0 and 1, with
higher values indicating stronger similarity between the two
solutions. Figure \ref{solutions} shows the similarities between
solutions obtained from applying the algorithm five different times
on the same network. For a given network, the $ij^{\text{th}}$ entry
in the lower triangle of the table is the Jaccard index for
solutions $i$ and $j$, while the $ij^{\text{th}}$ entry in the upper
triangle is the measure $f_{same}$ for solutions $i$ and $j$. We can
see that the solutions obtained from the five different runs are
similar, implying that the proposed label propagation algorithm can
effectively identify the community structure of any given network.
Moreover, the tight range and high values of the modularity measure
$Q$ obtained for the five solutions (figure \ref{solutions}) suggest
that the partitions denote significant community structures.

\subsection{Aggregate} 
\begin{figure*}
\begin{center}
  \includegraphics[width=10cm]{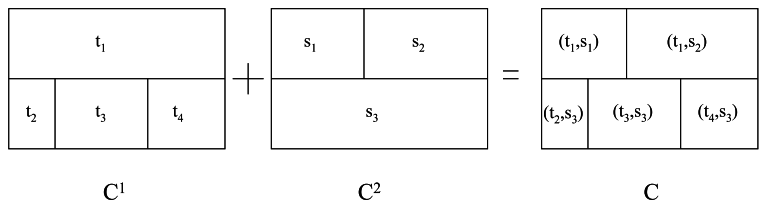}\\
  \caption{An example of aggregating two community structure solutions.
  $t_1,t_2,t_3$ and $t_4$ are labels on the nodes in a network obtained from
  solution 1 and denoted as $C^1$. The network is partitioned into groups
  of nodes having the same labels. $s_1,s_2$ and $s_3$ are labels on the nodes
  in the same network obtained from solution 2 and denoted as $C^2$. All nodes
  that had label $t_1$ in solution 1 are split into two groups with each group
  having labels $s_1$ and $s_2$ respectively. While all nodes with labels $t_3$
  or $t_4$ or $t_5$ in solution 1 have labels $s_3$ in solution 2. $C$ represents
  the new labels defined from $C^1$ and $C^2$.}\label{aggregate}
\end{center}
\end{figure*}

It is difficult to pick one solution as the
best among several different ones. Furthermore, one solution may be
able to identify a community that was not discovered in the other
and vice-versa. Hence an aggregate of all the different solutions
can provide a community structure containing the most useful
information. In our case a solution is a set of labels on the nodes
in the network and all nodes having the same label form a community.
Given two different solutions, we combine them as follows; let $C^1$
denote the labels on the nodes in solution 1 and $C^2$ denote the
labels on the nodes in solution 2. Then, for a given node $x$, we
define a new label as $C_x = (C^1_x,C^2_x)$ (see figure
\ref{aggregate}). Starting with a network initialized with labels
$C$ we perform the iterative process of label propagation until
every node in the network is in a community to which the maximum
number of its neighbors belong to. As and when new solutions are
available they are combined one by one with the aggregate solution
to form a new aggregate solution. Note that when we aggregate two
solutions, if a community $T$ in one solution is broken into two (or
more) different communities $S_1$ and $S_2$ in the other, then by
defining the new labels as described above we are showing
preferences to the smaller communities $S_1$ and $S_2$ over $T$.
This is only one of the many ways in which different solutions can
be aggregated. For other methods of aggregation used in community
detection refer to \cite{Wu04,Gfeller05,Wilkinson04}.

\begin{figure*}
\begin{center}
  \includegraphics[width=15cm]{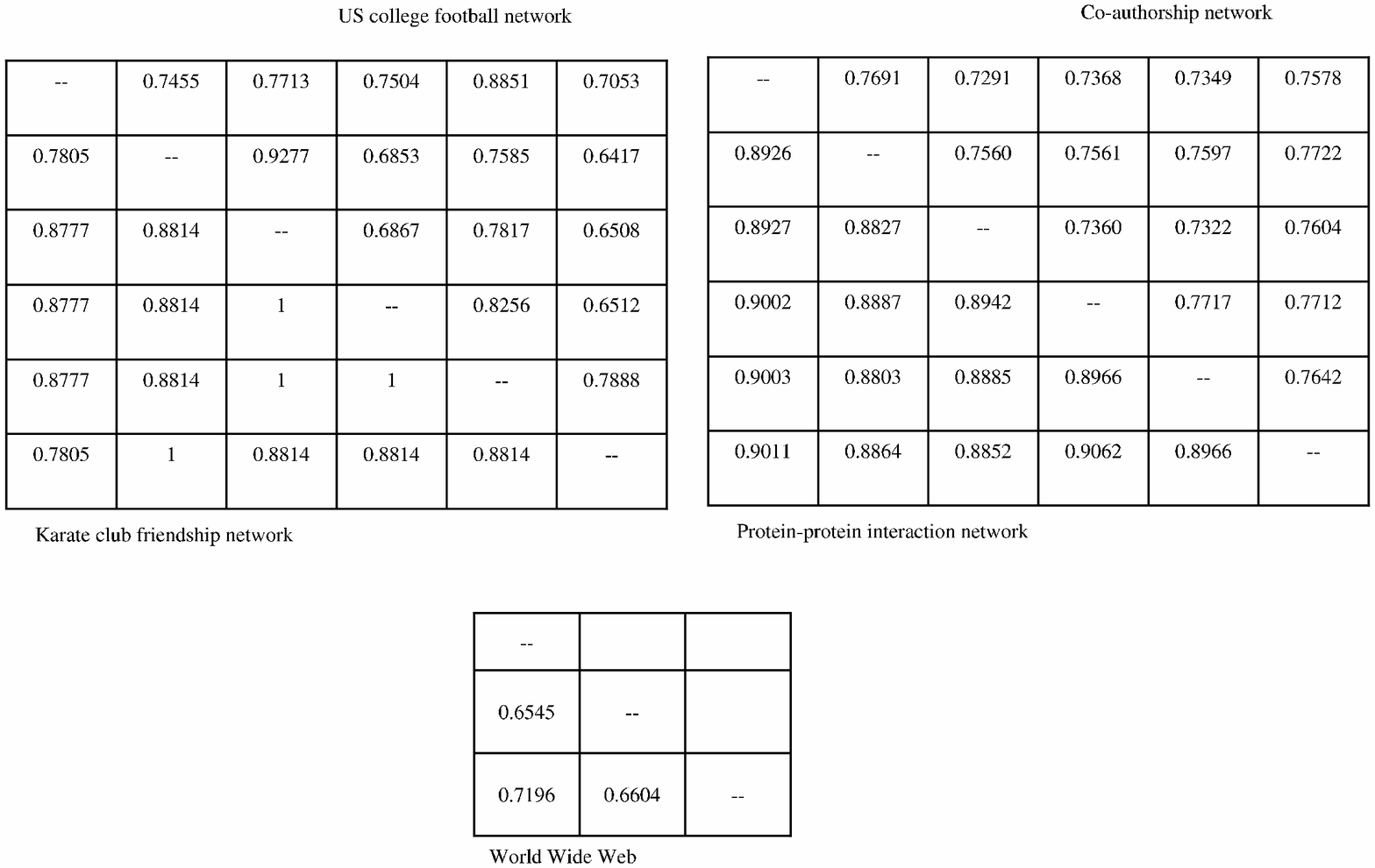}\\
  \caption{Similarities between aggregate solutions obtained for each network.
  An entry in the $i^{\text{th}}$ row and $j^{\text{th}}$ column
  in the tables is Jaccard's similarity index between the aggregate of
  the first $5i$ and the first $5j$ solutions. While similarities between solutions
  for the karate club friendship network and the protein-protein interaction network
  are represented in the lower triangles of the first two tables, the entries in the
  upper triangle of these two tables are for the US college football network and
  the co-authorship network respectively. The similarities between aggregate
  solutions for the WWW is given in the lower triangle of the third table.}\label{aggsolutions}
\end{center}
\end{figure*}

Figure \ref{aggsolutions} shows the similarities between aggregate
solutions. The algorithm was applied on each network 30 times and
the solutions were recorded. An $ij^{\text{th}}$ entry is the
Jaccard index for the aggregate of the first $5i$ solutions with the
aggregate of the first $5j$ solutions. We observe that the aggregate
solutions are very similar in nature and hence a small set of
solutions (5 in this case) can offer as much insight about the
community structure of a network as can a larger solution set. In
particular, the WWW network which had low similarities between
individual solutions (Jaccard index range 0.4883 - 0.5931), shows
considerably improved similarities (Jaccard index range 0.6604 -
0.7196) between aggregate solutions.

\section{Validation of the community detection algorithm}
Since we know the communities present in Zachary's karate club and
the US football network, we explicitly verify the accuracy of the
algorithm by applying it on these networks. We find that the
algorithm can effectively unearth the underlying community
structures in the respective networks. The community structures
obtained by using our algorithm on Zachary's karate club network is
shown in figure \ref{karatefootball}. While all the three solutions
are outcomes of the algorithm applied to the network, figure
\ref{karatefootball}$(b)$ reflects the true solution
\cite{Zachary77}.

Figure \ref{football} gives two solutions for the US college
football network. The algorithm was applied to this network 10
different times and the two solutions are the aggregate of the first
five and remaining five solutions. In both figures
\ref{football}$(a)$ and \ref{football}$(b)$, we can see that the
algorithm can effectively identify all the conferences with the
exception of Sunbelt. The reason for the discrepancy is the
following: among the seven teams in the Sunbelt conference, four
teams (Sunbelt$_4$ = \{North-Texas, Arkansas State, Idaho, New
Mexico State\}) have all played each other and three teams
(Sunbelt$_3$ = \{Louisiana-Monroe, Middle-Tennessee State,
Louisiana-Lafayette\}) have again played one another. {There is only
one game connecting} Sunbelt$_4$ and Sunbelt$_3$, {namely the game}
between North-Texas and Louisiana-Lafayette. However, four teams
from the Sunbelt conference (two each from Sunbelt$_4$ and
Sunbelt$_3$) have together played with seven different teams in the
Southeastern conference. Hence we have the Sunbelt conference
grouped together with the Southeastern conference in figure
\ref{football}$(a)$. In figure \ref{football}$(b)$, the Sunbelt
conference breaks into two, with Sunbelt$_3$ grouped together with
Southeastern and  Sunbelt$_4$ {grouped with an independent team
(Utah State), a team from Western Atlantic (Boise State), and the
Mountain West conference. The latter grouping is due to the fact
that every member of Sunbelt$_4$ has played with Utah State and with
Boise State, who have together played five games with four different
teams in Mountain West.} There are also five independent teams which
do not belong to any specific conference and are hence assigned by
the algorithm to a conference where they have played the maximum
number of their games.

\section{Time complexity}
It takes a near-linear time for the algorithm to run to its
completion. Initializing every node with unique labels requires
$O(n)$ time. Each iteration of the label propagation algorithm takes
linear time in the number of edges ($O(m)$). At each node $x$, we
first group the neighbors according to their labels ($O(d_x)$). We
then pick the group of maximum size and assign its label to $x$,
requiring a worst-case time of $O(d_x)$. This process is repeated at
all nodes and hence an overall time is $O(m)$ for each iteration.

As the number of iterations increases, the number of nodes that are
classified correctly increases. Here we assume that a node is
classified correctly if it has a label that the maximum number of
its neighbors have. From our experiments, we found that irrespective
of $n$, $95\%$ of the nodes or more are classified correctly by the
end of iteration 5. Even in the case of Erd\H{o}s - R\'enyi random
graphs \cite{Bollobas85} with $n$ between 100 and 10000 and average
degree 4, which do not have community structures, by iteration 5,
$95\%$ of the nodes or more are classified correctly. In this case,
the algorithm identified all nodes in the giant connected component
as belonging to one community.

When the algorithm terminates it is possible that two or more
disconnected groups of nodes have the same label (the groups are
connected in the network via other nodes of different labels). This
happens when two or more neighbors of a node receive its label and
pass the labels in different directions, which ultimately leads to
different communities adopting the same label. In such cases, after
the algorithm terminates one can run a simple breadth-first search
on the sub-networks of each individual groups to separate the
disconnected communities. This requires an overall time of $O(m+n)$.
When aggregating solutions however, we rarely find disconnected
groups within communities.

\section{Discussion and conclusions}
The proposed label propagation process uses only the network
structure to guide its progress and requires no external parameter
settings. Each node makes its own decision regarding the community
to which it belongs to based on the communities of its immediate
neighbors. These localized decisions lead to the emergence of
community structures in a given network. We verified the accuracy of
community structures found by the algorithm using Zachary's karate
club and the US college football networks. Furthermore, the
modularity measure $Q$ was significant for all the solutions
obtained, indicating the effectiveness of the algorithm. Each
iteration takes a linear time $O(m)$, and although one can observe
the algorithm beginning to converge significantly after about 5
iterations, the mathematical convergence is hard to prove. Other
algorithms that run in a similar time-scale include the algorithm of
Wu and Huberman \cite{Wu04} (with time complexity $O(m+n)$) and that
of Clauset et al \cite{Clauset04} which has a running time of $O(n
\text{log}^2n)$.

The algorithm of Wu and Huberman is used to break a given network
into only two communities. In this iterative process two chosen
nodes are initialized with scalar values 1 and 0 and every node
updates its value as the average of the values of its neighbors. At
convergence, if a maximum number of a node's neighbors have values
above a given threshold then so will the node. Hence a node tends to
be classified to a community to which the maximum number of its
neighbors belong. Similarly if in our algorithm we choose the same
two nodes and provide them with two distinct labels (leaving the
others unlabeled), the label propagation process will yield similar
communities as the Wu and Huberman algorithm. However to find more
than two communities in the network, the Wu and Huberman algorithm
needs to know a priori how many communities there are in the
network. Furthermore, if one knows that there are $c$ communities in
the network, the algorithm proposed by Wu and Huberman can only find
communities that are approximately of the same size, that is
$\frac{n}{c}$, and it is not possible to find communities with
heterogeneous sizes. The main advantage of our proposed label
propagation algorithm over the Wu and Huberman algorithm is that we
do not need a priori information on the number and sizes of the
communities in a given network; indeed such information usually is
not available for real-world networks. Also, our algorithm does not
make restrictions on the community sizes. It determines such
information about the communities by using the network structure
alone.

\begin{figure*}
\begin{center}
  \includegraphics[width=10.8cm]{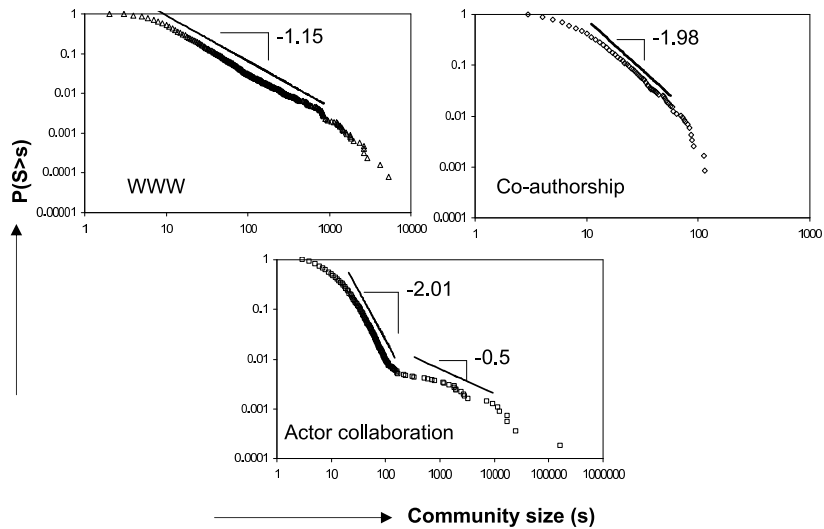}\\
  \caption{The cumulative probability distributions of community sizes ($s$) are
  shown for the WWW, co-authorship and actor collaboration networks. They approximately follow
  power-laws with the exponents as shown.}\label{sizedistribution}
\end{center}
\end{figure*}

In our test networks, the label propagation algorithm found
communities whose sizes follow approximately a power-law
distribution {$P(S>s) \sim s^{-\nu}$} with the {exponent} $\nu$
ranging between 0.5 and 2 {(figure \ref{sizedistribution})} . This
implies that there is no characteristic community size in the
networks and it is consistent with previous observations
\cite{Arenas04,Clauset04,Newman04b}. While the community size
distributions for the WWW and co-authorship networks approximately
follow power-laws with a cut-off, with exponents 1.15 and 1.98
respectively, there is a clear crossover from one scaling relation
to another for the actor collaboration network. The community size
distribution for the actor collaboration network has a power-law
exponent of 2 for sizes up to 164 nodes and 0.5 between 164 and 7425
nodes (see figure \ref{sizedistribution}).

In the hierarchical agglomerative algorithm of Clauset et al
\cite{Clauset04}, the partition that corresponds to the maximum $Q$
is taken to be the most indicative of the community structure in the
network. Other partitions with high $Q$ values will have a structure
similar to that of the maximum $Q$ partition, as these solutions are
obtained by progressively aggregating two groups at a time. Our
proposed label propagation algorithm on the other hand finds
multiple significantly modular solutions that have some amount of
dissimilarity. For the WWW network in particular, the similarity
between five different solutions is low, with the Jaccard index
ranging between 0.4883 to 0.5921, yet all five are significantly
modular with $Q$ between 0.857 to 0.864. This implies that the
proposed algorithm can find not just one but multiple significant
community structures, supporting the existence of overlapping
communities in many real-world networks \cite{Palla05}.

\section{Acknowledgments}
The authors would like to acknowledge the National Science
Foundation (SST 0427840, DMI 0537992 and CCF 0643529) and a Sloan
Research Fellowship to one of the authors (R. A). Any opinions,
findings and conclusions or recommendations expressed in this
material are those of the author(s) and do not necessarily reflect
the views of the National Science Foundation.

\bibliographystyle{apsrev}
\bibliography{Biblio-Database}

\end{document}